\documentstyle{article}
\begin{document}

\title{\Large\bf
THE GENERATIONAL SEESAW MECHANISM IN [SU(6)]$^{3}\times$Z$_{3}$}
\author{\large
William A. Ponce$^{1,2}$, Arnulfo Zepeda$^2$ \\
and Ricardo Gaitan Lozano$^{2,3}$ \\
\normalsize  1-Departamento de F\'\i sica, Universidad de Antioquia\\
\normalsize A.A. 1226, Medell\'\i n, Colombia \\
\normalsize  2-Departamento de F\'\i sica, Centro de Investigaci\'on y
de Estudios Avanzados del I.P.N. \\
\normalsize Apartado Postal 14-740, 07000 M\'exico D.F., M\'exico \\
\normalsize 3-Departamento de F\'\i sica, Universidad Surcolombiana \\
\normalsize A.A. 385, Neiva, Colombia}

\renewcommand{\baselinestretch}{1.2}
\setlength{\hsize}{15cm}

\date{August 15, 1992}
\maketitle

\large

\newcommand{\suty}{SU(3)$_{c}\otimes$U(1)$_{Y_{(B-L)}}$}
\newcommand{\lefty}{SU(2)$_{L}\otimes$SU(2)$_{R}\otimes$U(1)$_{Y_{(B-L)}}$}
\newcommand{\colorem}{SU(3)$_{c}\otimes$U(1)$_{EM}$\hspace{0.2cm}}
\newcommand{\sm}{SU(3)$_{c}\otimes$SU(2)$_{L}\otimes$U(1)$_{Y}$}

\vspace{2cm}

\begin{center}
{\bf ABSTRACT}
\end{center}
\large
\parbox{15cm}%
{In the context of the left-right symmetric gauge group [SU(6)]$^3\times$
Z$_3$ which unifies nongravitational forces with flavors, we analyze the
generational seesaw mechanism. At tree level we get
m$_{\nu_\tau}\sim$m$_{\nu_\mu}\sim$M$^2_L$/M$_H$,
m$_{\nu_e}=0$, where M$_L\sim10^2$ GeVs and M$_H\ge 100$ TeVs is the
mass scale at which the horizontal interactions get spontaneously broken.
The right handed neutrinos get a Majorana mass M$_R\gg $M$_H$ of the order of
the scale where SU(2)$_R$ is broken. An exotic neutral lepton with a mass of
the order of M$^2_H$/M$_R$ is predicted. First order radiative corrections
will produce m$_{\nu_e}\neq0$ which is
at least two orders of magnitude smaller than the other two neutrino masses.}
\vspace{.5cm}

----------------------------------------------------------\\
\small E-mail: azepeda@cinvesmx
\pagebreak

\large

\large
\section{Introduction}
In Ref. \cite{wp} to \cite{pz} we have presented
a model based upon the gauge group
G $\equiv$ [SU(6)]$^{3}\times$Z$_{3}$ which unifies
the known nongravitational interactions with flavors. Our aim
is to discuss the fermion mass hierarchy problem. Since G includes the so
called horizontal interactions, it leads to some predictions for the
masses of some elementary fermions.

As it is clear from Refs. \cite{wp} to \cite{pz}, in our model the three
families of ordinary fermions belong to
a single irreducible representation (irrep) of G and therefore to three
identical irreps of
SU(3)$_c\otimes$SU(2)$_L\otimes$SU(2)$_R\otimes$U(1)$_{(B-L)}$, the
left-right symmetric extension of the standard model (LRSM).

The seesaw mechanism\cite{seesaw}
expresses the smallness of the neutrino
masses in terms of the ``large" masses of some other neutral fermions.
This mechanism is very simple to implement for a single neutrino and, as
far as we know, it has been implemented in a consistent way only for
two families\cite{dw}, with a convincing generational seesaw mechanism for
three families still lacking.  In this paper we show how the
generational seesaw mechanism emerges naturally in the context of
G. By naturally we mean that it arises as a by-product of
spontaneously breaking the symmetry G down to \colorem
with what we have called the minimal set of Higgs Fields (HF) and
Vacuum Expectation Values (VEVs).

The rest of the paper is as follows: In section 2 we briefly
review the model; in section 3 we show how the generational
seesaw mechanism arises in the context of our model; in the last
section we present our conclusions and show how the one loop radiative
corrections slightly modify our main results.

\vspace {3mm}
\section{The model}
SU(6)$_{L}\otimes$SU(6)$_{C}\otimes$SU(6)$_{R}\times$Z$_{3}$
is our  gauge group. SU(6)$_C$ is a vector like
group which includes three hadronic and three leptonic
colors. It includes as a subgroup the
SU(3)$_c\otimes$U(1)$_{(B-L)}$ group of the LRSM.
SU(6)$_{L}\otimes$SU(6)$_{R}$ is
the left-right symmetric flavor group which includes the
SU(2)$_{L}\otimes$SU(2)$_{R}$ gauge group of the LRSM and
a SU(3)$_{HL}\otimes$SU(3)$_{HR}$  horizontal gauge group.

The 105 gauge bosons and the 108 Weyl fermions in G are clearly
defined in \cite{pz}. Just for the sake of what follows, let us
write here some of the Weyl fermions:
The ordinary (known) fermions are included in: \\
$\psi(108)_L =\psi(6,1,\bar{6})_{L} +
\psi(1,\bar{6},6)_{L} + \psi(\bar{6},6,1)_{L}$,
which has the following particle content:

\begin{equation}
\psi(\bar{6},6,1)_{L} = \left( \begin{array}{cccccc}
{\rm d}_{x}^{-1/3} & {\rm d}_{y}^{-1/3} & {\rm d}_{z}^{-1/3} & E_{1}^{-} &
L_{1}^{0} & T_{1}^{-}\\
{\rm u}_{x}^{2/3} & {\rm u}_{y}^{2/3} & {\rm u}_{z}^{2/3} & E_{1}^{0} &
L_{1}^{+} & T_{1}^{0}\\
{\rm s}_{x}^{-1/3} & {\rm s}_{y}^{-1/3} & {\rm s}_{z}^{-1/3} & E_{2}^{-} &
L_{2}^{0} & T_{2}^{-}\\
{\rm c}_{x}^{2/3} & {\rm c}_{y}^{2/3} & {\rm c}_{z}^{2/3} & E_{2}^{0} &
L_{2}^{+} & T_{2}^{0}\\
{\rm b}_{x}^{-1/3} & {\rm b}_{y}^{-1/3} & {\rm b}_{z}^{-1/3} & E_{3}^{-} &
L_{3}^{0} & T_{3}^{-}\\
{\rm t}_{x}^{2/3} & {\rm t}_{y}^{2/3} & {\rm t}_{z}^{2/3} & E_{3}^{0} &
L_{3}^{+} & T_{3}^{0}
\end{array}
\right) _{L}\equiv\psi^{\alpha}_{a},
\end{equation}

\noindent
where the rows (columns) represent color (flavor) degrees of
freedom; E$_i^{-,0}$, L$_i^{+,0}$, and T$_i^{-,0}$ $i=1,2,3$ stand
for leptonic fields with electrical charges as indicated, and
d,u,s,c,b,and t stand for the corresponding quark fields
eigenstates of G. The
$x,y,z$ subindices in the quark fields are SU(3)$_c$ color
indices.

$\psi(1,\bar{6},6)_L\equiv\psi^A_\alpha$ stands for  the 36
fields charge conjugate to the right handed parts of the
fields in $\psi(\bar{6},6,1)_L$, while
$\psi(6,1,\bar{6})_{L}$ represents  36 exotic Weyl leptons: 9 with
positive electric charges, 9 with negative (the charge
conjugated to the positive ones) and 18 are
neutrals.
As it is clear we are using a,b,... (A,B,...) [$\alpha, \beta,...$]
as SU(6)$_L$  (SU(6)$_R$) [SU(6)$_C$] tensor indices.

The most economical set of HF and VEVs which breaks the symmetry
from G down to \colorem and at the same time gives  tree level
mass only to the top quark and seesaw masses to the
tau and mu neutrinos (what we call
the {\bf modified horizontal survival hypothesis} \cite{gmpz}) is
\cite{pz}:

\begin{equation}
\phi_{1}=\phi(675)=
\phi_{1,[a,b]}^{[A,B]}+\phi^{[\alpha,\beta]}_{1,[A,B]}
+\phi^{[a,b]}_{1,[\alpha,\beta]}       \label{eq: fi1}
\end{equation}

\noindent
with VEVs in the directions [a,b]=[1,6]=$-$[2,5]=$-$[3,4], [A,B] similar to
[a,b] and [$\alpha,\beta$]=[5,6];

\begin{equation}
\phi_{2}=\phi(1323)=
\phi_{2,\{a,b\}}^{\{A,B\}}+\phi^{\{\alpha,\beta\}}_{2,\{A,B\}}
+\phi^{\{a,b\}}_{2,\{\alpha,\beta\}}       \label{eq: fi2}
\end{equation}

\noindent
with VEVs in the directions \{a,b\}=\{1,4\}=$-$\{2,3\}, \{A,B\} similar to
\{a,b\} and \{$\alpha,\beta$\}=\{4,5\};

\begin{equation}
\phi_{3}=\phi^\prime(675)=
\phi_{3,[a,b]}^{\prime,[A,B]}+\phi^{\prime,[\alpha,\beta]}_{3,[A,B]}
+\phi^{\prime,[a,b]}_{3,[\alpha,\beta]}       \label{eq: fi3}
\end{equation}

\noindent
with VEVs such that $\langle\phi_{3,[a,b]}^{\prime,[A,B]}\rangle=
\langle\phi^{\prime,[a,b]}_{3,[\alpha,\beta]}\rangle=0$, and
$\langle\phi^{\prime,[\alpha,\beta]=[4,6]}_{3,[A,B]=[4,6]}\rangle=$M$_R$;

\begin{equation}
\phi_4=\phi(108)=\phi^A_{4,\alpha}+\phi^\alpha_{4,a}+\phi^a_{4,A}.
\label{eq: fi4}
\end{equation}

\noindent
with VEVs such that
$\langle\phi_\alpha^A\rangle=\langle\phi_a^\alpha\rangle=0$ and
$\langle\phi_A^a\rangle=$M$_L$, with
values different from zero only in the directions
$\langle\phi_2^2\rangle=
\langle\phi_4^2\rangle=
\langle\phi_6^2\rangle=
\langle\phi_2^4\rangle=
\langle\phi_4^4\rangle=
\langle\phi_6^4\rangle=
\langle\phi_2^6\rangle=
\langle\phi_4^6\rangle=
\langle\phi_6^6\rangle=$M$_{L}$.

In Eqs. (~\ref{eq: fi1} -  ~\ref{eq: fi3}) \{.,.\}
and [.,.] stand respectively for the symmetrization and
antisymmetrization
of the indices inside the brackets. The non-zero components of
$\langle\phi_1\rangle$ and $\langle\phi_2\rangle$
are set equal to M and M$^\prime$ respectively. Since at this stage
we are not interested in studying CP violation,
we will assume throughout the paper that
$\langle\phi_i\rangle$, $i=1,2,3,4$, are real numbers.

According to the analysis presented in Ref. \cite{pz},
$\langle\phi_1+\phi_2\rangle$ breaks G down to the LRSM
group; $\langle\phi_3\rangle$ alone breaks G down to a rather
complicated gauge structure (see appendix A in Ref. \cite{pz}), but
$\langle\phi_1+\phi_2+\phi_3\rangle$ breaks G down to \sm, the  gauge
group of the Standard Model (SM).

\section{Generational Seesaw Mechanism}
The HF and VEVs presented in the previous
section imply the following Yukawa-type mass terms:

\begin{eqnarray}
\psi^\alpha_a\psi^\beta_b\langle
\phi^{[a,b]}_{1,[\alpha,\beta]}
+\phi^{\{a,b\}}_{2,\{\alpha,\beta\}}
\rangle+\psi_\alpha^A\psi_\beta^B\langle
\phi_{1,[A,B]}^{[\alpha,\beta]}
+\phi_{2,\{A,B\}}^{\{\alpha,\beta\}}
+\phi_{3,[A,B]}^{\prime,[\alpha,\beta]}\rangle \nonumber \\
+\psi_A^a\psi_B^b\langle
\phi_{1,[a,b]}^{[A,B]}
+\phi_{2,\{a,b\}}^{\{A,B\}}\rangle
+\sum_{\alpha,a,A=1}^6\psi^\alpha_a\psi^A_\alpha\langle\phi^a_A\rangle
+ h.c.
\end{eqnarray}
\noindent
The tree level mass matrices for the quarks and the charged leptons produced
by this expression are quite simple to analyze and that task
was done already in Ref. \cite{pz}. In order to diagonalize the
mass matrix for
the neutral leptons, let us write it first in the basis defined by
$\stackrel{\rightarrow}{{\bf N_0}}=\\
\noindent
(E_1^0$, $E_2^0$, $E_3^0$, $T_1^0$, $T_2^0$, $T_3^0$, $L_1^{0c}$,
$L_2^{0c}$, $L_3^{0c}$,
$E_1^{0c}$, $E_2^{0c}$, $E_3^{0c}$, $T_1^{0c}$, $T_2^{0c}$,
$T_3^{0c}$, $L_1^0$, $L_2^0$, $L_3^0)_L$,

\noindent
where the upper $c$ symbol denotes the fields
in $\psi(1,\bar{6},6)_{L}$. In this basis
the mass matrix has the form:

\pagebreak
\small

\[
{\cal M}_{tree} =
\]

\renewcommand{\baselinestretch}{1.5}

\[
\left(\begin{array}{cccccccccccccccccc}
0 & 0 & 0 & 0 & 0 & 0 & 0 & 0 & 0 &
{\rm M}_L & {\rm M}_L & {\rm M}_L & 0 & 0 & 0 & 0 & -{\rm M}^\prime & 0  \\
0 & 0 & 0 & 0 & 0 & 0 & 0 & 0 & 0 &
{\rm M}_L & {\rm M}_L & {\rm M}_L & 0 & 0 & 0 & {\rm M}^\prime & 0 & 0 \\
0 & 0 & 0 & 0 & 0 & 0 & 0 & 0 & 0 &
{\rm M}_L & {\rm M}_L & {\rm M}_L & 0 & 0 & 0 & 0 & 0 & 0 \\
0 & 0 & 0 & 0 & 0 & 0 & 0 & 0 & 0 &
0 & 0 & 0 & {\rm M}_L & {\rm M}_L & {\rm M}_L & 0 & 0 & {\rm M}  \\
0 & 0 & 0 & 0 & 0 & 0 & 0 & 0 & 0 &
0 & 0 & 0 & {\rm M}_L & {\rm M}_L & {\rm M}_L & 0 & -{\rm M} & 0 \\
0 & 0 & 0 & 0 & 0 & 0 & 0 & 0 & 0 &
0 & 0 & 0 & {\rm M}_L & {\rm M}_L & {\rm M}_L & {\rm M} & 0 & 0 \\
0 & 0 & 0 & 0 & 0 & 0 & 0 & 0 & 0 &
0 & {\rm M}^\prime & 0 & 0 & 0 & {\rm M} & 0 & 0 & 0  \\
0 & 0 & 0 & 0 & 0 & 0 & 0 & 0 & 0 &
-{\rm M}^\prime & 0 & 0 & 0 & -{\rm M} & 0 & 0 & 0 & 0 \\
0 & 0 & 0 & 0 & 0 & 0 & 0 & 0 & 0 &
0 & 0 & 0 & {\rm M} & 0 & 0 & 0 & 0 & 0 \\
{\rm M}_L & {\rm M}_L & {\rm M}_L & 0 & 0 & 0 & 0 & -{\rm M}^\prime & 0  &
0 & 0 & 0 & 0 & 0 & 0 & 0 & 0 & 0  \\
{\rm M}_L & {\rm M}_L & {\rm M}_L & 0 & 0 & 0 & {\rm M}^\prime & 0 & 0 &
0 & 0 & 0 & 0 & 0 & {\rm M}_R & 0 & 0 & 0 \\
{\rm M}_L & {\rm M}_L & {\rm M}_L & 0 & 0 & 0 & 0 & 0 & 0 &
0 & 0 & 0 & 0 & -{\rm M}_R & 0 & 0 & 0 & 0 \\
0 & 0 & 0 & {\rm M}_L & {\rm M}_L & {\rm M}_L & 0 & 0 & {\rm M}  &
0 & 0 & 0 & 0 & 0 & 0 & 0 & 0 & 0 \\
0 & 0 & 0 & {\rm M}_L & {\rm M}_L & {\rm M}_L & 0 & -{\rm M} & 0 &
0 & 0 & -{\rm M}_R & 0 & 0 & 0 & 0 & 0 & 0 \\
0 & 0 & 0 & {\rm M}_L & {\rm M}_L & {\rm M}_L & {\rm M} & 0 & 0 &
0 & {\rm M}_R & 0 & 0 & 0 & 0 & 0 & 0 & 0 \\
0 & {\rm M}^\prime & 0 & 0 & 0 & {\rm M} & 0 & 0 & 0  &
0 & 0 & 0 & 0 & 0 & 0 & 0 & 0 & 0  \\
-{\rm M}^\prime & 0 & 0 & 0 & -{\rm M} & 0 & 0 & 0 & 0 &
0 & 0 & 0 & 0 & 0 & 0 & 0 & 0 & 0  \\
0 & 0 & 0 & {\rm M} & 0 & 0 & 0 & 0 & 0 &
0 & 0 & 0 & 0 & 0 & 0 & 0 & 0 & 0
\end{array} \right) \nonumber \label{eq: mx1}
\]

\begin{flushright}
(7)
\end{flushright}

\pagebreak

\renewcommand{\baselinestretch}{1.2}

\large

\noindent
where
M$_L\sim10^2$ GeVs as usual because it is the mass parameter which
characterizes the breaking of SU(2)$_L$.
On the other hand, since the gauge
bosons responsible for the horizontal transitions in the model get
masses of order\cite{gmpz,pz} M and M$^\prime$, and since the horizontal
transitions include flavor changing neutral currents, we must
impose the experimental \cite{fcnc} constrain M,M$^\prime>100$ TeVs.
And about M$_{R}$, it is
a common prejudice that, since it is the mass scale for
the right-handed neutrinos, and since in most of the models it
is responsible for the seesaw mechanism, then
M$_R\sim 10^{11,12}$ GeVs. Because of these arguments, it is
natural to think that we can diagonalize {\cal M}
under the assumption that M$_R\gg$M$\sim$M$^\prime\gg$M$_L$ by the
use of a double perturbation theory. In what follows we are going to assume
such a mass hierarchy.

To diagonalize ${\cal M}_{tree}$ we first identify the neutrino states
(right and
left) for the particular case M$_L=$M$_R=0$ for which ${\cal M}_{tree}$ has
rank twelve.
The six zero eigenvalues are associated with the
following eigenvectors: \\
i-(E$^{0}_3$; (ME$^{0}_2-$M$^\prime$T$^{0}_3$)/$V$;
(ME$^{0}_1-$M$^\prime$T$^{0}_2$)/$V)_L$, with
$V=($M$^2+$M$^{\prime,2})^{1/2}$, which we define as
($\nu_1,\nu_2,\nu_3)_L$, due to the fact that its components
constitute a basis for the
physical neutrinos $\nu_e,\nu_\mu,\nu_\tau$  (they are
SU(2)$_L$ doublets and SU(2)$_R$ singlets); \\
ii-(E$^{0c}_3$; (ME$^{0c}_2-$M$^\prime$T$^{0c}_3$)/$V$;
(ME$^{0c}_1-$M$^\prime$T$^{0c}_2$)/$V$)$_L$ which we define as
($\nu_1^c,\nu_2^c,\nu_3^c)_L$ , due to the fact that its components
constitute a basis for the right-handed neutrinos
$\nu_e^c,\nu_\mu^c,\nu_\tau^c$ (they are SU(2)$_L$ singlets and
SU(2)$_R$ doublets).

This suggest to use, instead of the basis
$\stackrel{\rightarrow}{{\bf N_0}}$, a more convenient one defined by

\noindent
$\stackrel{\rightarrow}{{\bf N_1}}=(\nu_1$, $\nu_2$, $\nu_3$, N$_1$, N$_2$,
N$_3$, L$_1^c$, L$_2^c$, L$_3^c$, $\nu_1^c$, $\nu_2^c$, $\nu_3^c$,
N$_1^c$, N$_2^c$, N$_3^c$, L$_1$, L$_2$, L$_3)_L$,

\noindent
where $N_1=T_1^0$, $N_2=($M$T_3^0+$M$^\prime E_2^0)/V$,
$N_3=($M$T^0_2+$M$^\prime E_1^0)/V$, $L_i=L_i^0,$ $i=1,2,3$,
$N_1^c=T_1^{0c}$, etc.

\subsection{First perturbation}
{}From now on let M=M$^\prime$=M$_H$. In the first approximation
given by M$_L=0$ and in the basis
$\stackrel{\rightarrow}{{\bf N_1}}$ we then have

\pagebreak
\small
\[
{\cal M}_{1}^{2} =
\]

\renewcommand{\baselinestretch}{1.5}

\[
\left(\begin{array}{cccccccccccccccccc}
0 & 0 & 0 & 0 & 0 & 0 & 0 & 0 & 0 & 0 & 0 & 0 & 0 & 0 & 0 & 0 & 0 & 0  \\
0 & 0 & 0 & 0 & 0 & 0 & 0 & 0 & 0 & 0 & 0 & 0 & 0 & 0 & 0 & 0 & 0 & 0  \\
0 & 0 & 0 & 0 & 0 & 0 & 0 & 0 & 0 & 0 & 0 & 0 & 0 & 0 & 0 & 0 & 0 & 0  \\
0 & 0 & 0 & {\rm M}^2_H & 0 & 0 & 0 & 0 & 0 & 0 & 0 & 0 & 0 & 0 & 0 & 0 & 0 & 0
    \\
0 & 0 & 0 & 0 & 2{\rm M}^2_H & 0 & 0 & 0 & 0 & 0 & 0 & 0 & 0 & 0 & 0 & 0 & 0 &
0
     \\
0 & 0 & 0 & 0 & 0 & 2{\rm M}^2_H & 0 & 0 & 0 & 0 & 0 & 0 & 0 & 0 & 0 & 0 & 0 &
0
     \\
0 & 0 & 0 & 0 & 0 & 0 & 2{\rm M}^2_H & 0 & 0 & 0 & 0 & 0 & 0 & \sqrt2 {\rm
M}_H{
   \rm M}_R & 0 & 0 & 0 & 0 \\
0 & 0 & 0 & 0 & 0 & 0 & 0 & 2{\rm M}^2_H & 0 & {\rm M}_H{\rm M}_R & 0 & 0 & 0 &
   0 & 0 & 0 & 0 & 0  \\
0 & 0 & 0 & 0 & 0 & 0 & 0 & 0 & {\rm M}^2_H & 0 & 0 & 0 & 0 & 0 & 0 & 0 & 0 & 0
    \\
0 & 0 & 0 & 0 & 0 & 0 & 0 & {\rm M}_H{\rm M}_R & 0 & {\rm M}_R^2 & 0 & 0 & 0 &
0
    & 0 & 0 & 0 & 0 \\
0 & 0 & 0 & 0 & 0 & 0 & 0 & 0 & 0 & 0 & {\rm M}_R^2 & 0 & 0 & 0 & 0 & 0 & 0 & 0
   \\
0 & 0 & 0 & 0 & 0 & 0 & 0 & 0 & 0 & 0 & 0 & {\rm M}_R^2/ \sqrt2 & 0 & 0 & -{\rm
   M}_R^2/2 & 0 & 0 & 0 \\
0 & 0 & 0 & 0 & 0 & 0 & 0 & 0 & 0 & 0 & 0 & 0 & {\rm M}^2_H & 0 & 0 & 0 & 0 & 0
    \\
0 & 0 & 0 & 0 & 0 & 0 & \sqrt2 {\rm M}_H{\rm M}_R & 0 & 0 & 0 & 0 & 0 & 0 & d_1
   & 0 & 0 & 0 & 0  \\
0 & 0 & 0 & 0 & 0 & 0 & 0 & 0 & 0 & 0 & 0 & -{\rm M}_R^2/2 & 0 & 0 & d_2 & 0 &
0
    & 0 \\
0 & 0 & 0 & 0 & 0 & 0 & 0 & 0 & 0 & 0 & 0 & 0 & 0 & 0 & 0 & 2{\rm M}^2_H & 0 &
0
     \\
0 & 0 & 0 & 0 & 0 & 0 & 0 & 0 & 0 & 0 & 0 & 0 & 0 & 0 & 0 & 0 & 2{\rm M}^2_H &
0
     \\
0 & 0 & 0 & 0 & 0 & 0 & 0 & 0 & 0 & 0 & 0 & 0 & 0 & 0 & 0 & 0 & 0 & {\rm M}^2_H
    \\
\end{array} \right)  \nonumber \label{eq: mx2}
\],

\begin{flushright}
(8)
\end{flushright}

\pagebreak

\renewcommand{\baselinestretch}{1.2}
\large

\noindent
where
$d_1=$M$_R^2$+2M$^2_H$ and  $d_2=$M$_R^2/2$+2M$^2_H$.
This matrix has rank fifteen,
with the three eigenvalues equal to zero belonging to the
three left-handed neutrinos.
We  diagonalize it perturbatively under the
assumption that M$_R\gg$M$_H$.
The list of eigenvalues and eigenvectors of ${\cal M}^2_1$ is,
up to second order in perturbation theory, the following
(where our expansion parameter is $\delta =$M/M$_R$):

\begin{itemize}
\item $\nu _1, \nu _2$, and $\nu _3$ have eigenvalue zero;
\item $N_1, L_3^c, N_1^c$, and $L_3$ have eigenvalue M$^2_H$;
\item $N_2, N_3, L_1$, and $L_2$ have eigenvalue $V^2=2$M$^2_H$;
\item $\nu _2^c$ has eigenvalue M$_R^2$;
\item \mbox{$\sqrt{2}(\delta -3\delta^3)L_1^c+(1-\delta^2)N_2^c
\equiv L_{1s}$ has eigenvalue M$_R^2(1+4\delta^2)\equiv \alpha^2_1$;}
\item $(1-\delta^2)L_1^c-\sqrt{2}(\delta -3\delta^3)N_2^c
\equiv N_{2s}$ has eigenvalue 2M$_R^2\delta^4\equiv\alpha^2_2$;
\item \mbox{$\frac{1}{\sqrt{2}}(1-\delta ^2)\nu ^c_3-\frac{1}{\sqrt{2}}
(1+\delta ^2)N_3^c\equiv\nu _{3s}$ has eigenvalue
M$_ R^2(1+\delta ^2)\equiv\beta ^2_1$;}
\item $\frac{1}{\sqrt{2}}(1+\delta ^2)\nu ^c_3+\frac{1}{\sqrt{2}}
(1-\delta ^2)N_3^c\equiv N_{3s}$ has eigenvalue M$_R^2\delta ^2\equiv\beta
^2_2$
   .
\item \mbox{$(\delta +\delta ^3/2)L_2^c+(1-\delta ^2/2)\nu _1^c\equiv L_{2s}$
has eigenvalue M$_R^2(1+\delta ^2)=\beta ^2_1$;}
\item $(1-\delta ^2/2)L^c_2-(\delta+\delta ^3/2)\nu ^c_1\equiv\nu _{1s}$
has eigenvalue M$_R^2\delta ^2=\beta ^2_2$.
\end{itemize}

These eigenvectors define a new basis which we denote by
$\stackrel{\rightarrow}{{\bf N_2}}$.
Notice that $L_{1s},\nu _{3s}$ and $L_{2s}$ have
eigenvalues of order M$_R^2$; while $N_{3s}$ and $\nu _{1s}$ have eigenvalues
of order M$^2_H$, but $N_{2s}$ has an eigenvalue of order M$^4/$M$_R^2$ which
is a seesaw eigenvalue produced by the two mass scales M$_H$ and M$_R$. This
state has important consequences for low energy phenomenology because it is
the only low energy exotic predicted by this model.

\subsection{Second perturbation}
$\stackrel{\rightarrow}{{\bf N_2}}$ diagonalizes ${\cal M}_1^2$.
To diagonalize ${\cal M}_1$ we apply an orthogonal transformation to
$\stackrel{\rightarrow}{{\bf N_2}}$
and obtain
$\stackrel{\rightarrow}{{\bf N_2^\prime}}$
given by

\noindent
$\stackrel{\rightarrow}{{\bf N_2^\prime}}=(\nu_1$, $\nu_2$, $\nu_3$,
N$^\prime_1
   $,
N$^\prime_2$, N$^\prime _3$, L$_1^{\prime ,c}$, L$_2^{\prime ,c}$,
L$_3^{\prime ,c}$, $\nu_1^{\prime ,c}$, $\nu_2^c$,
$\nu_3^{\prime ,c}$, N$_1^{\prime ,c}$, N$_2^{\prime ,c}$,
N$_3^{\prime ,c}$, L$_1^\prime$, L$_2^\prime$, L$_3^\prime)_L$,

\noindent
where $\nu _i,i=1,2,3$ and $\nu _2^c$ are the same as in
$\stackrel{\rightarrow}{{\bf N_1}}$, but
N$^\prime _1= $(N$_1+$L$_3)/\sqrt{2}$,
N$^\prime _2= $(N$_2+$L$_1)/\sqrt{2}$,
N$^\prime _3= $(N$_3+$L$_2)/\sqrt{2}$,
L$^\prime _1= $(N$_1-$L$_3)/\sqrt{2}$,
L$^\prime _2= $(N$_3-$L$_2)/\sqrt{2}$,
L$^\prime _3= $(N$_2-$L$_1)/\sqrt{2}$,
L$_1^{\prime ,c}$=L$_{1s}$,
L$_2^{\prime ,c}= $(L$_{2s}+\nu_{3s})/\sqrt{2}$,
L$_3^{\prime ,c}= $(L$_3^{c}+$N$_1^c)/\sqrt{2}$,
$\nu _1^{\prime ,c}= $(L$_{2s}-\nu_{3s})/\sqrt{2}$,
$\nu_3^{\prime ,c}= $(N$_{3s}+\nu_{1s})/\sqrt{2}$,
N$_1^{\prime ,c}= $(L$_{3}^{c}-$N$_1^{c})/\sqrt{2}$,
N$_2^{\prime c}=$N$_{2s}$ and
N$_3^{\prime ,c}= $(N$_{3s}-\nu_{1s})/\sqrt{2}$.

\noindent
In this new basis the mass matrix ${\cal M}_{tree}$ can be written as:

\begin{equation}
{\cal M}={\cal M}_D+{\cal V}_m
\end{equation}

\noindent
where ${\cal M}_D$ is a diagonal $18\times 18$ mass matrix with entries:\\
${\cal M}_D$=Diag.(0,0,0,M$_H$,$\sqrt2$ M$_H$,-$\sqrt2$ M$_H$,
$\alpha_1,\beta_1$,M$_H$,-$\beta_1$,-M$_R$,-$\beta_2$,-M$_H$,
-$\alpha_2$,$\beta_2$,-M$_H$,$\sqrt2$ M$_H$,-$\sqrt2$ M$_H$),
(where all the symbols stand for absolute values. Notice
that $tr{\cal M}_D=0$ as it should be according to Eq. (7) and
${\cal V}_m$ given below).
${\cal V}_m$ is a perturbation which can be
written as:

\begin{equation}
\begin{array}{ccc}
{\cal V}_m=\left( \begin{array}{ccc}
0_{6\times 6} & A_{6\times 9} & 0_{6\times 3}  \\
A_{9\times 6} & 0_{9\times 9} & B_{9\times 3}  \\
0_{3\times 6} & B_{3\times 9} & 0_{3\times 3} \end{array} \right), &
A_{6\times 9}=\left( \begin{array}{c}
A_{3\times 9} \\
B_{3\times 9}  \end{array}  \right),  &
A_{9\times 6}=\left( \begin{array}{cc}
A_{9\times 3} & B_{9\times 3} \end{array} \right),
\end{array}
\end{equation}

\noindent
where $0_{n\times m}$ are zero matrices with $n$ rows and $m$ columns, and
$A_{3\times 9}=A^T_{9\times 3}$ and $B_{3\times 9}=B^T_{9\times 3}$ are
given by:

\begin{equation}
A_{3\times 9}=\frac{{\rm M}_L}{\sqrt{2}}\left(
\begin{array}{ccccccccc}
\kappa _1 & \eta _1 & 0 & \eta _2 & 1 & \eta _3 & 0 & \kappa _2 & \eta _4 \\
0 & \eta_5 & -1/\sqrt{2} & \eta _6 & \sqrt{2} &
\eta _7 & 1/\sqrt{2} & 0 & \eta _8 \\
0 & \eta_5 & -1/\sqrt{2} & \eta _6 & \sqrt{2} &
\eta _7 & 1/\sqrt{2} & 0 & \eta _8
\end{array} \right)  \label{eq: A39},
\end{equation}

\begin{equation}
B_{3\times 9}=\frac{{\rm M}_L}{2}\left(
\begin{array}{ccccccccc}
\kappa _1 & -1 & 1 & 1 &-1 & -\delta^2 & -1 & \kappa_2 & -\delta^2  \\
\sqrt{2}\kappa_1 & -\delta^2 & 1/\sqrt{2} & \sqrt{2} & 0 & \rho_1 &
-1/\sqrt{2} & \sqrt{2}\kappa_2 & \rho_2\\
\sqrt{2}\kappa_1 & -\delta^2 & 1/\sqrt{2} & \sqrt{2} & 0 & \rho_1 &
-1/\sqrt{2} & \sqrt{2}\kappa_2 & \rho_2
\end{array} \right),
\end{equation}

\noindent
respectively. In these last two matrices,
 $\kappa _1=1-\delta ^2$, $\kappa _2=-\sqrt{2}\delta(1-3\delta
^2)$,$\eta_1=1-3\delta^2/2$, $\eta_2=1+\delta^2/2$,
$\eta_3=1-\delta-\delta^3/2$
   ,
$\eta_4=-(1+\delta+\delta^3/2)$, $\eta_5=\sqrt{2}(1-3\delta^2/4)$,
$\eta_6=\delta^2/2\sqrt{2}$, $\eta_7=(1-\delta+\delta^2-\delta^3/2)/\sqrt{2}$,
$\eta_8=-(1+\delta+\delta^2+\delta^3/2)/\sqrt{2}$,
$\rho_1=(1-\delta-\delta^2-\delta^3/2)/\sqrt{2}$, and
$\rho_2=-(1+\delta-\delta^2+\delta^3/2)/\sqrt{2}$.

As we will see in what follows, the most important thing about the former
algebraic results is the particular form which $A_{3\times 9}$ takes. As
stated, ${\cal V}_m$ is a perturbation to ${\cal M}_D$ and it produces
corrections to the eigenvalues and eigenvectors of ${\cal M}_D$ of the
order of M$_L/$M$\equiv\xi$, M$_L/$M$_R\equiv\xi^\prime$
and smaller (higher orders). These corrections
are important only for the smaller eigenvalues, i.e., for the eigenvalues
corresponding to $\nu_1,\nu_2,\nu_3$ and $N_2^{\prime,c}$. For these states
we use matrix perturbation theory\cite{landau}
in order to get the corrections.

The second order perturbative corrections
to the $3\times 3$ mass matrix for the states $(\nu_1,\nu_2,\nu_3)$ show up
after diagonalizing the following mass matrix:

\begin{equation}
C_{n,n^\prime}=\sum_{m=1}^{9}\frac{(A_{3\times 9})_{n,m}
(A_{9\times 3})_{m,n^\prime}}{({\cal M}_D)_{m+3,m+3}}.
\end{equation}

\noindent
Then, according to matrix Eq. (~\ref{eq: A39}) $C$ has the following form:

\begin{equation}
C=\left( \begin{array}{ccc}
\theta_3 & \theta_1 & \theta_1 \\
\theta_1 & \theta_2 & \theta_2  \\
\theta_1 & \theta_2 & \theta_2
\end{array} \right)
\end{equation}

\noindent
where $\theta_1=-$M$_L(\xi+3\delta\xi^\prime)\sqrt{2}$,
$\theta_2=$M$_L\xi^\prim
   e$,
and $\theta_3=6$M$_L\delta^2\xi^\prime\sim0$.

Now, independent of the value for $\theta_3$(zero or not), $C$ is a
rank two matrix. The two eigenvalues different from zero are approximately
given by M$_L(\xi+3\delta\xi^\prime\pm\xi^\prime)\simeq
$M$_L\xi$=M$^2_L$/M$_H$.

 The state associated with $N_2^{0,c}$ is not degenerate, so a
straightforward second order perturbation calculation gives a mass
correction = M$_L\xi^\prime/\sqrt{2}$=M$^2_L$/$\sqrt{2}$ M$^2_R$ which
is smaller than its original value (M$^2_H$/M$_R$).

\section{Conclusions}
In the context of the model presented in Ref. \cite{wp} to \cite{pz} and
for the mass
hierarchy M$_R\gg$ M$_H \gg$ M$_L$ we have diagonalized the $18\times 18$
mass matrix for the neutral leptons. The original mass matrix includes only
tree level mass terms and the diagonalization was done by using a double
perturbation  theory with corrections up to second order in the parameters.

Our analysis gave four neutral leptons with small masses. Three
of them are mainly members of
left handed doublets, one with zero mass and two with seesaw masses of order
M$_L^2$/M$_H$. The fourth is mainly member of a right handed doublet, left
handed singlet, with
a seesaw mass of order M$^2_H$/M$_R$.

If we make the assumption that the neutrinos do not oscillate, then  we can
identify the real neutrino states as the mass eigenstates
(neutrino oscillations can be analyzed in the context of our model, but it is
a tougher matter because it requires to identify simultaneously  the known
charged lepton states, which in turn requires a consistent treatment of the
mass
radiative corrections). We therefore obtain at tree level

\phantom{yyyyyyyyyyyyyyyyyyy}m$_{\nu_e}=0$,

m$_{\nu_\mu}$=m$_{\nu_\tau}\simeq$ M$_L^2$/M$_H$.

\noindent
These results and the experimental upper limit \cite{massnumu}
 m$_{\nu_\mu}<0.27$ MeVs imply M$_H >10^7$
GeVs,  which is consistent (although discouraging) with the experimental
constraint \cite{fcnc}  M$_H >10^5$ GeVs. Also the experimental lower
limit of
10 GeVs for the mass of any exotic neutral lepton imposes, via the result
for the $N_2^{\prime c}$ mass value, the relation M$_R$<M$^2_H$/10 GeV.

When the first order mass radiative corrections are included, then we expect
modification in all our former results of the order of $\delta^2$/M$_H$, where
$\delta \sim 1$ GeV. is the order of the radiative masses expected in our
model (the radiative corrections must produce masses for all the known
particles but the t quark, according to the modified horizontal survival
hypothesis). Then we should expect m$_{\nu_e} \sim \delta ^2$/M$_H$, at least
tw
   o
or three orders of magnitude smaller than the other two neutrino masses.

Now, looking our algebraic results, we realize immediately that the common
prejudice of using the value M$_R\sim 10^{12}$ GeVs from the seesaw mechanism
for the neutrinos, is not well founded , because with the above assummed
hierarchy it is the
intermediate mass scale M$_H$ the responsible for the seesaw mechanism.
If we repeat the calculations for the mass
hierarchy M$_H \gg $M$_R \gg $M$_L$, then we get results very similar to
the previous ones with the roles of M$_H$ and M$_R$ interchanged. That is,
it is now the intermediate mass scale M$_R$ the responsible for the
seesaw mechanism. The fact is that, for any mass hierarchy, and due to the
particular form of the mass matrix ${\cal M}_{tree}$ (democratic in flavor
for the entries proportional to M$_L$), both mass scales M and M$_R$
produces seesaw mechanisms, but obviously, it is the lower mass scale,
that is the  nearest neighbord, the
one which produces the dominant effect.

\noindent
{\bf ACKNOWLEDGMENTS}
\noindent
This work was partially supported by CONACyT in M\'exico and COLCIENCIAS in
Colombia. One of us benefited from the fruitful atmosphere of the Aspen Center
for Physics during the course of this work. A discussion with R. Shrock
in relation with FCNC limits is acknowledged.


\begin{thebibliography}{99}

\bibitem{wp}
W.A. Ponce, in {\it Particles and Fields}, Proceedings of the Third
Mexican School of Particles and Fields, Oaxtepec, Mexico, 1988. Ed. by
J.L. Lucio and A. Zepeda (World Scientific, Singapore, 1989, pp. 90-129);
W.A. Ponce and A. Zepeda, CINVESTAV report, 1991 (unpublished).

\bibitem{gmpz}
A.H. Galeana, R. Martinez, W.A. Ponce, and A. Zepeda; Phys. Rev.
{\bf D44}, 2166(1991).

\bibitem{pz}
W.A. Ponce, and A. Zepeda  ; {\it ``Tuning} [SU(6)]$^3\times$Z$_3$".
CINVESTAV preprint(1992). Submitted for publication.

\bibitem{seesaw}
M. Gell-Mann, P. Ramond and R. Slansky, in {\it ``Supergravity"}, eds.
D. Freedman and P. van Nieuwenhuizen (North-Holland, Amsterdam, 1979).\\
T. Yanahida, in {\it ``Proceedings of the workshop on unified theories
and Baryon number in the Universe"}, eds. A. Sawada and A. Sugamoto
(KEK Report No. 79-18, Tsukuba, Japan, 1979).

\bibitem{dw}
A.D. Davidson, and K.C. Wali; Phys. Rev. Lett. {\bf 60}, 1813 (1988).

\bibitem{fcnc}
T. Akagi {\it et al.}; Phys. Rev. Letters {\bf 67} 2614 (1991);\\
A.M. Lee {\it et al.}; Phys. Rev. Letters {\bf 64} 165 (1990);\\
R. Scrock, private communication.


\bibitem{landau}
L.D. Landau, and E.M. Lifshitz; {\it ``Quantum Mechanics, Nonrelativistic
Theory"}, Addison Wesley P.C., Reading MA. 1958, first published
English edition,. p.p.137-140.

\bibitem{massnumu}
R. Abela {\it et al.}; Phys. Lett. {\bf 146B} 431 (1984);\\
B. Jeckelman {\it et al.}; Pyes. Rev. Letters {\bf 56} 1444 (1986)

\end{thebibliography}
\end{document}